\documentclass[pdftex,twocolumn,epjc3]{svjour3}          % twocolumn

\RequirePackage[T1]{fontenc}

\smartqed  % flush right qed marks, e.g. at end of proof

\usepackage{amsmath}
\usepackage{subcaption}
\usepackage{color}
\RequirePackage{graphicx}
\RequirePackage[colorlinks,citecolor=blue,urlcolor=blue,linkcolor=blue]{hyperref}

\journalname{Eur. Phys. J. C}

\begin{document}

\title{Quantum Larmor radiation in de Sitter spacetime}

\author{Robert Blaga\thanksref{e1,addr1}
%etc.
        \and
        Sergiu Busuioc\thanksref{e2,addr1}
}

\thankstext{e1}{e-mail: robert.blaga@e-uvt.ro}
\thankstext{e2}{e-mail: sergiu.busuioc@e-uvt.ro}

\institute{West University of Timi\c soara,
V.  P\^ arvan Ave.  4, RO-300223 Timi\c soara, Romania\label{addr1}
}
\date{Received: date / Accepted: date}

\maketitle

\begin{abstract}
We study the radiation emitted by inertial charge evolving on the expanding de Sitter spacetime. Performing a perturbative calculation, within scalar quantum electrodynamics (sQED), we obtain the transition amplitude for the process and using this we define the energy radiated by the source. In the non-relativistic limit we find that the leading term is compatible with the classical result (Larmor formula). The first quantum correction is found to be negative, a result which is in line with a number of similar quantum field theory results. For the ultra-relativistic case we find a logarithmic divergence of the emitted energy for large frequencies, which we link to the nature of the spacetime.  We compare our results with that of Nomura et al. (2006), where the authors make a similar calculation for a general conformally flat spacetime.
\end{abstract}
\section{Introduction}
It is a well known result in classical electrodynamics that accelerated charges radiate. The emitted power is given by the famous Larmor formula \cite{Jackson}. The radiated energy in the case of non-relativistic motion of the source and with acceleration parallel to the velocity, adjusted for units, can be written as
\begin{eqnarray} \label{Larmor}
E_{cl}  &=& \frac{e^2}{6\pi} \int \ddot{x}(t)^2 dt \quad
\end{eqnarray}
It is expected that the same result can be recovered from quantum theory in the limit $\hbar \rightarrow 0$. Indeed, in Ref.\cite{Higuchi}, the authors obtained from sQED the lowest order contribution as being in agreement with the Larmor formula. The authors considered two distinct cases of external electromagnetic fields that give rise to the same classical acceleration. Interestingly, although the leading term in both cases agrees with the classical result, the main quantum corrections differ. Similar results were obtained in Refs.\cite{N3,N4} in a spatially homogeneous time-dependent electric field and electromagnetic plane-wave background. \par
A distinct problem is the radiation of a charge in a time-dependent spacetime. In this case, in the GR picture, the source is inertial and the dynamic background plays the role of the external field. The problem was tackled in Refs.\cite{N1,N2}, in the general case of a conformally flat spacetime, by using the WKB approximation for the mode functions. The authors found that the leading term reproduces exactly the relativistic version of (\ref{Larmor}), when the trajectory is expressed in terms of conformal time. \par
Because of its privileged position in cosmological physics,  the case of the de Sitter spacetime deserves a separate, more detailed treatment. Our goal in this paper is to  obtain the term corresponding to the classical radiation and to calculate the leading quantum corrections for the energy radiated by a charge evolving on the expanding de Sitter spacetime (dS). We approach the problem with a perturbative calculation within sQED. We derive the radiated energy from the 1st order transition amplitude of the process which is analogous to the classical one. \par
One might wonder why there is radiation at all, given that the source is inertial (i.e. follows a geodesic trajectory). The motion of charges in gravitational fields has produced some controversy over the past decades, resulting in a considerable amount of literature on the subject \cite{deWitt,GronGrav,Poisson}. The peculiarities of the problem are nicely illustrated by Chiao's paradox \cite{Chiao,Crowell}. The question asked by Chiao is the following: will a charge on a circular orbit around a planet radiate and thus spiral inwards, as Newtonian intuition predicts, or continue moving along the geodesic, in accordance with the equivalence principle\,? The paradox can be solved by noting that the equivalence principle has only local validity, while an electromagnetic charge along with its field is an extended object.  "The Coulomb field of the particle, as it sweeps over the 'bumps' in spacetime, receives 'jolts' that are propagated back to the particle. [...] The radiated effect comes from the work performed by this force."\cite{GronRev} An important feature of this radiation is that it is observer dependent. The classical example is that of the uniformly accelerated charge in flat space. While an inertial observer sees the charge radiating according to the Larmor formula, a co-accelerated observer will detect no radiation.\cite{Rohrlich} A similar situation arises in de Sitter space for comoving versus non-comoving observers. On physical grounds we expect that similarly to the uniformly accelerated case \cite{GronRev}, the radiation reaction on a charge in dS cancels out, leaving the particle on the initial (geodesic) trajectory. The rule of thumb is: if there is variation in the local (physical) momentum of the charge in the relative motion with respect to the observer, there will be radiation. \par
The paper is structured as follows: In sec.II we gather the basics of sQED on de Sitter spacetime. In sec.III, starting from the transition amplitude, we define the energy radiated through the process. We obtain an asymptotic form for the energy in a weak gravitational field and proceed by expanding the result for different regimes of  motion of the source. In the non-relativistic limit, we obtain the leading term and the first quantum contributions to the radiated energy. We obtain also a closed form for the energy in the ultra-relativistic limit. We find that the total emitted energy in this case is plagued by divergences, which is a typical feature of dS. In sec.IV we summarize and discuss our main results. \par
 We work in natural units where $\hbar = c = 1$.
\section{Basics of sQED on dS}
The expanding patch of the de Sitter spacetime is described by the line element
\begin{equation} \label{dS}
ds^2 = dt^2 - \mathrm{e}^{2\omega t} d\vec{x}^2 = \frac{1}{(\omega\eta)^2}(d\eta^2 - d\vec{x}^2),
\end{equation}
where $\omega$ is the Hubble constant, and we have introduced for convenience the time parameter $\omega\eta= e^{-\omega t}, \ \eta \in (0,\infty)$, with opposite sign as compared to what one usually calls conformal time. \par
The scalar modes that define the Bunch-Davies vacuum are \cite{Birrell,C1}
\begin{equation} \label{BD modes}
f_{\vec{p}}(x) = \frac{1}{2}\sqrt{\frac{\pi}{\omega}} \frac{(\omega\eta)^{3/2}}{(2\pi)^{3/2}}\mathrm{e}^{i\pi\nu/2}  \mathcal{H}_\nu^{(1)}\left(p\eta\right) e^{i\vec{p}\vec{x}},
\end{equation}
where $\nu = i\sqrt{\mu^2 - 9/4}$, and $m = \omega\mu$ is the mass of the scalar field. The momenta appearing in (\ref{BD modes}) are the conformal momenta which are related to the physical momenta as $p=\frac{p_{\text{phys}}}{\omega\eta}$, and $p^0 = \sqrt{p^2 + \left(\frac{m}{\omega\eta}\right)^2}$.\par
Given that the electromagnetic field is conformally invariant, the covariant components of the field are identical to their Minkowskian counterparts, while the contravariant components can be obtained by raising the indices with the metric tensor. The mode functions for the Maxwell field are \footnote{Notice that the sign in $\mathrm{e}^{ik\eta}$ is positive. This is due to the fact that the conformal time is equal to $(-\eta)$ .}\cite{C3}:
\begin{equation} \label {EM modes}
w^i_{\vec{k},\lambda}(x) = \frac{1}{(2\pi)^{3/2}}\frac{1}{\sqrt{2k}}\mathrm{e}^{ik\eta + i\vec{k}\vec{x}} \varepsilon^i_\lambda(\vec{k})\mathrm{e}^{-2\omega t}.
\end{equation}
\par
Following Ref.\cite{C4} we work with the electromagnetic field in the Coulomb gauge, given by $A^0 = 0,$ \\$\left(\sqrt{-g}A^i\right)_{;i} = 0$. \par
Given the mode functions (\ref{BD modes}) and (\ref{EM modes}), we can write the usual mode expansion for the scalar and Maxwell fields as:
\begin{eqnarray}
\varphi(x) &=& \int d^3p \left(\,a(\vec{p})f_{\vec{p}}(x) + b^\dag(\vec{p})f^*_{\vec{p}}(x)\,\right) \\
A^i(x) &=& \sum_\lambda \int d^3k \left(\, c_\lambda(\vec{k})w^i_{\vec{k},\lambda}(x) + c^\dag_\lambda(\vec{k})w^{i*}_{\vec{k},\lambda}(x)\right), \nonumber
\end{eqnarray}
where $a^\dag(\vec{p}),b^\dag(\vec{p}),c^\dag(\vec{k})$ are the respective creation operators for the scalar particles, antiparticles and photons. \par
We are interested here only in tree-level QED processes, that are generated by the 1st order term in the expansion of the S-matrix:
\begin{equation}
S^{(1)} = e\int \sqrt{-g}\left( \varphi^\dag(x) \stackrel{\leftrightarrow}{\partial}_\alpha \varphi(x) \right) A^\alpha(x).
\end{equation}
We have dropped the four-point interaction term because it does not contribute to the process studied here. \\
We use a general prescription for interacting fields, following the classical textbook \cite{Birrell}.
A detailed treatment of sQED on dS, including the reduction mechanism, can be found in Ref.\cite{C2}. \par
The transition amplitude can be written as:
\begin{equation}
\mathcal{A}_{in  \rightarrow out} = \langle out \vert\, S^{(1)}\, \vert in \rangle
\end{equation}
\section{Quantum radiation of scalar charges}
%\section{Energy emission through QED process}
\begin{figure*}
\centering
\begin{subfigure}[b]{0.5\linewidth}
\centering
\includegraphics[width=3in]{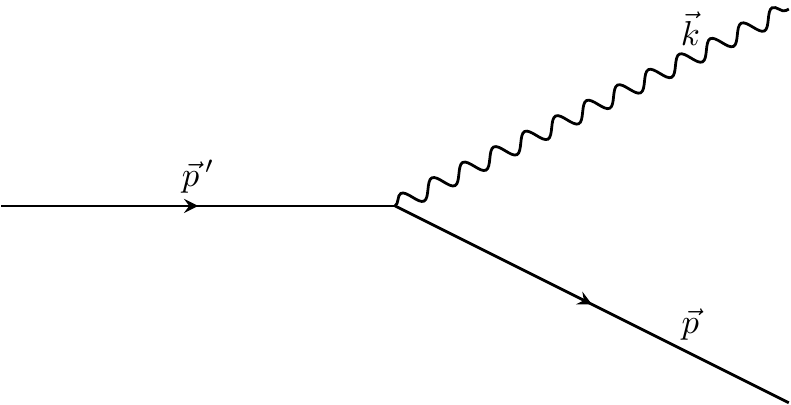}
\caption{}
\label{1a}
\end{subfigure}%
\begin{subfigure}[b]{0.5\linewidth}
\centering
\includegraphics[width=3in]{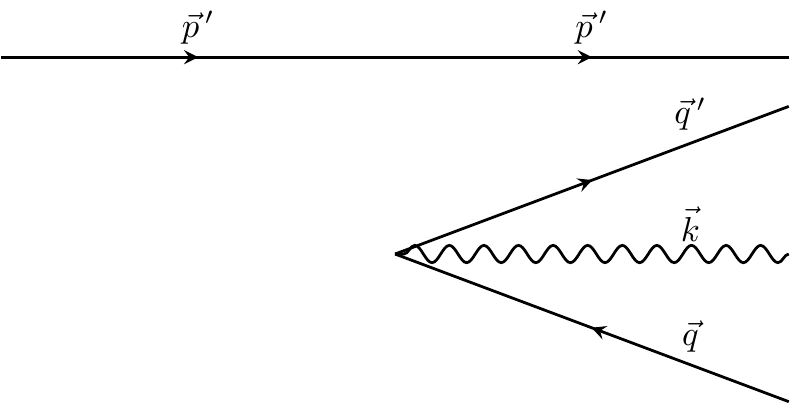}
\caption{}
\label{1b}
\end{subfigure}
\caption{\small Feynman diagram for (a) photon emission and (b) triplet production. }
 \label{1}
\end{figure*}
We are interested in the quantum theoretical counterpart of a charge emitting electromagnetic radiation given an external influence. In our case the charge is inertial, the expanding background playing the role of the external influence. The setup is as follows: in the initial state there is one scalar particle with momentum $p'$, and in the final state we have a photon with momentum $k$ and an arbitrary state of the scalar field. We average over all configurations that are indistinguishable from the point of view of a detector measuring the emitted radiation. This means basically summing over all possible final states of the scalar field. \par
\begin{eqnarray} \label{SumProb}
E &\sim& \sum_{a,b^*}\, \left\vert  \langle 1_{\vec{k},\lambda};\, a_\varphi, b^*_{\varphi^\dag} \,\vert S^{(1)}\vert\ 1_{\vec{p}'} \rangle \right\vert^2 \\
&=& \sum_{a,b^*} \, \langle 1_{\vec{p}'} \vert S^{(1)*}\vert\ 1_{\vec{k},\lambda};\, a_\varphi, b^*_{\varphi^\dag}\, \rangle\langle 1_{\vec{k},\lambda};\, a_\varphi, b^*_{\varphi^\dag}\,\vert S^{(1)} \vert\ 1_{\vec{p}'} \rangle, \nonumber
\end{eqnarray}
where $a$ and $b^*$ represent the number of scalar particles and antiparticles. Notice that the quantity (\ref{SumProb}) is independent of the definition of particles in the $out$ state. Indeed this is the case because we can factor out an identity in (\ref{SumProb}), which can in turn be replaced by any complete orthonormal basis. The most natural way to proceed is in fact to insert an $in$ basis (built from the Bunch-Davies modes (\ref{BD modes})), which then truncates the sum at a finite number of terms. In our case we are left with the following terms:
\begin{align} \label{Emission+Prod}
E \sim \left\vert \langle 1_{\vec{k},\lambda}; 1_{\vec{p}} \,\vert S^{(1)} \vert 1_{\vec{p}'} \rangle \right\vert^2 + \left\vert  \langle 1_{\vec{k},\lambda};\, 1_{\vec{p}'},1_{\vec{q}'}, 1^*_{\vec{\vec{q}}} \,\vert S^{(1)}\vert 1_{\vec{p}'} \rangle \right\vert^2
\end{align}
The first term represents a particle emitting a photon, while the second term represents the particle passing through without interacting, accompanied by the production of a pair and a photon from the vacuum\footnote{Note that momentum conservation constrains the momenta \\ $\vec{p}+\vec{k} =\vec{p}'$ in the first, and $\vec{k}+\vec{q}+\vec{q}'=0$ in the second process.}. The two configurations are illustrated by the Feynman diagrams Fig.\ref{1} . A very important observation is that the second process yields homogeneous and isotropic radiation. In an experimental context we can imagine that the detector can be adjusted to account for this background radiation. We can then drop this contribution and focus on the first term only.

\subsection{Transition amplitude and radiated energy}
The amplitude corresponding to the process depicted in Fig.\ref{1a} was obtained in ref.\cite{B2} by one of the authors, and equals:
 \begin{eqnarray} \label{Amp}
\mathcal{A}(\vec{p}',\vec{p},\vec{k}) &=& \langle 1_{\vec{k},\lambda};\, 1_{\vec{p}} \,\vert\ 1_{\vec{p}'} \rangle \\
&=& -e\int \mathrm{d}^4x \sqrt{-g}\left( f^*_{\vec{p}}(x)\stackrel{\leftrightarrow}{\partial}_i f_{\vec{p}'}(x) \right) w^{i\,*}_{\vec{k},\lambda}(x)\nonumber\\
%&&\left. \qquad- \partial_i f^*_{\vec{p}}(x) f_{\vec{p}'}(x) \right) w^{i*}_{\vec{k},\lambda}(x) \nonumber\\
&=& -\delta^3(\vec{p'}-\vec{p}-\vec{k})\frac{ie\pi(\vec{p'} + \vec{p})\cdot \vec{\varepsilon}^*_\lambda(\vec{k})}{4(2\pi)^{3/2}\sqrt{2k}} \nonumber\\
 &&\times  \int_0^\infty d\eta\, \eta \mathcal{H}_\nu^{(1)}(p'\eta)\mathcal{H}_\nu^{(2)}(p\eta)\,e^{-ik\eta-\epsilon\eta} \nonumber
\end{eqnarray}
In Ref.\cite{B2} we numerically analysed the result of the temporal integration and found a closed analytical form for the amplitude. Here instead we are interested only in the weak gravitational field limit ($m/\omega\rightarrow\infty$). With this in mind, we search for an asymptotic expression of the amplitude, in order to obtain the emitted energy as a power series in the Hubble constant $\omega$. \par
The energy emitted through the process can be computed as the energy of a photon $\hbar k$, weighted with the probability of emitting a photon with the corresponding momentum. The expression for the energy can be written as:
\begin{eqnarray} \label{energia}
E &=&   \sum_{\lambda} \frac{(2\pi)^3}{V}\int d^3k\int d^3p\ \hbar k\, \left\vert \mathcal{A}(\vec{p}',\vec{p},\vec{k}) \right\vert ^2 \\
&\equiv& \int  \frac{dE}{dk\,d\Omega}\,dk d\Omega \quad \nonumber,
\end{eqnarray}
where $V$ is the conformal volume, that will cancel the $\delta(0)$ term from the amplitude via the usual trick.
Making use of momentum conservation $\vec{p}' = \vec{p}+\vec{k}$, the polarization term in (\ref{Amp}) gives:
\begin{eqnarray} \label{polarizare}
\sum_\lambda \left\vert (\vec{p}' + \vec{p})\cdot \vec{\varepsilon}^*_\lambda(\vec{k}) \right\vert^2 &=& 4\left( {\vec{p}'}^2 - \frac{(\vec{p}'\cdot\vec{k})^2}{k^2}\right) \\
&=& 4{p'}^2\sin^2\theta. \nonumber
\end{eqnarray}

Integrating over the final momentum of the source and with the use of (\ref{polarizare}), the radiated energy becomes:
\begin{eqnarray} \label{dEdk}
E &=& \int d^3k\ \frac{e^2\pi^2}{16}\frac{4{p'}^2\sin^2\theta}{2(2\pi)^3} \\
&& \times  \left\vert\int_0^\infty d\eta\, \eta \,\mathcal{H}_\nu^{(1)}(p'\eta)\mathcal{H}_\nu^{(2)}(p\eta)\,e^{-ik\eta-\epsilon\eta} \right\vert^2. \nonumber
\end{eqnarray}
\par
In what follows we attempt to find an asymptotic form for the temporal integral in (\ref{dEdk}).
\subsection{Asymptotic expression in weak gravitational field}
We seek an asymptotic expression for the radiated energy in the weak gravitational field regime. The idea is to obtain the energy as a series in powers of the Hubble constant $\omega$. The leading term should be independent of $\hbar$, so that we can consider it the "classical" radiation, i.e. it should reproduce the result obtained from classical electrodynamics. Our expectation is enforced by the results obtained in ref.\cite{N1} for a conformally flat universe (of which dS is a particular instance of). The calculation in \cite{N1} was performed in the WKB approximation, and the condition for weak gravitational field ($ \mu \rightarrow \infty$) indeed assures that the WKB condition is fulfilled in our case also.
%We thus expect to reproduce the results of refs.\ref{japonezi, etc} quite accurately. \par

To obtain an asymptotic expansion of (\ref{Amp}) we start by writing the Hankel functions $\mathcal{H}_\nu^{(1)},\mathcal{H}_\nu^{(2)}$ , in terms of modified Bessel functions $K_\nu$ \cite{AS}:
\begin{align}
  \mathcal{H}_\nu^{(1)}(ze^{\frac{ i\pi}{2}}) &= \frac{2}{i\pi}e^{-\frac{i\pi\nu}{2}}K_\nu(z) \\
  \mathcal{H}_\nu^{(2)}(ze^{\frac{ -i\pi}{2}}) &= -\frac{2}{i\pi} e^{\frac{i\pi\nu}{2}}K_\nu(z). \nonumber
\end{align}
Using the property of the modified Bessel functions
\begin{equation}
K_\nu(z)=K_{-\nu}(z),
\end{equation}
we write the product of Hankel functions as follows:
\begin{align}\label{notationhh}
I_{p,p'}^\nu(\eta)&=\mathcal{H}_\nu^{(1)}(p'\eta)\mathcal{H}_\nu^{(2)}(p\eta)\\
&=\frac{4}{\pi^2} K_{-\nu}\left(p'\eta e^{-\frac{i\pi}{2}}\right) K_{\nu}\left(p\eta e^{\frac{i\pi}{2}}\right). \nonumber
\end{align}
Next, we use a large argument expansion \cite{AS}:
\begin{equation}\label{mf}
K_\nu(\nu z) \simeq \sqrt{\frac{\pi}{2\nu}}\frac{e^{-\nu \xi}}{\sqrt[4]{1 + z^2}}\Big\{1 + \sum_{k = 1}^{\infty}(-)^k\frac{u_k(t)}{\nu^k}\Big\},
\end{equation}
where:
\begin{align}
&\xi = \sqrt{1 + z^2} + \ln{\frac{z}{1 + \sqrt{1 + z^2}}} \nonumber\\
&u_1(t) = \frac{3t - 5t^3}{24}, \ t = \frac{1}{\sqrt{1 + z^2}}, \nonumber
\end{align}
which holds uniformly for $|\arg{z}|<\frac{1}{2}\pi$ when $\nu \rightarrow \infty$\footnote{We numerically tested that (\ref{mf}) holds also for complex indices i.e. for $|\nu|\rightarrow \infty$.}. The sign of the indices in (\ref{notationhh}) has been taken so that the condition on $\arg{z}$ is always fulfilled.

Substituting the expansion (\ref{mf}) into (\ref{notationhh}) and keeping only terms up to order $\mathcal{O}(\frac{1}{\mu^2})$, we obtain:
\begin{align}\label{masterformula}
  I_{p,p'}^\nu(\eta) &= \frac{2}{\pi\mu}\frac{e^{i\mu\xi'}}{\sqrt[4]{1 + {z'}^{2}}}\frac{e^{-i\mu\xi}}{\sqrt[4]{1 + z^2}}\\
  &\times\left( 1 + \frac{1}{i\mu}\frac{3t'-5t'^3}{24}\right)\left( 1 - \frac{1}{i\mu}\frac{3t-5t^3}{24}\right) \nonumber \\
    &= \frac{2}{\pi\mu} \frac{e^{i\mu\sqrt{1 + {z'}^{2}}}}{\sqrt[4]{1 + {z'}^{2}}}\frac{e^{-i\mu\sqrt{1+z^2}}}{\sqrt[4]{1 + z^2}} \nonumber\\
    &\times \left(\frac{z'}{1 + \sqrt{1 + {z'}^{2}}}\right)^{i\mu} \Bigg(\frac{z}{1 + \sqrt{1 + z^2}}\Bigg)^{-i\mu} \nonumber\\
    &\times\left( 1 + \frac{1}{i\mu}\frac{3t'-5t'^3}{24}\right) \left( 1 - \frac{1}{i\mu}\frac{3t-5t^3}{24}\right), \nonumber
\end{align}
where $z' = \frac{p'\eta}{\mu}$, $z = \frac{p\eta}{\mu}$ and we have considered $\nu\simeq i\mu$.\par

The temporal integral with the expansion (\ref{masterformula}) can not be solved analytically. To continue, we need to further expand the asymptotic formula for small and large values of z. By observing that
\begin{equation}
z = \frac{p\eta}{\mu} = \frac{p_{phys}}{m},
\end{equation}
we can properly consider $p_{phys}\ll m$  to be a non-relativistic approximation, while $p_{phys}\gg m$ represents an ultra-relativistic limit.

\subsection{Radiation in the non-relativistic limit}

First we discuss the radiation in the non-relativistic limit ($z \ll 1$). Expanding all functions around small $z$ and again keeping terms only up to order $\mathcal{O}\left(\frac{1}{\mu^2}\right)$, the asymptotic expression (\ref{masterformula}) reduces to:
\begin{align}\label{hh}
\nonumber I_{p,p'}^\nu(\eta) &\simeq \frac{2}{\pi\mu}\frac{e^{i\mu(1 + \frac{1}{2}{z'}^2)}}{(1 + \frac{1}{4}{z'}^2)}\frac{e^{-i\mu(1 + \frac{1}{2}z^2)}}{(1 + \frac{1}{4}{z}^2)}\left(\frac{\,p'}{p}\right)^{i\mu} \\
\nonumber &\times\left(\frac{2 + \frac{1}{2}{z}^2}{2 + \frac{1}{2}{z'}^2}\right)^{-i\mu} \Bigg(1 - \frac{1}{12i\mu}\Bigg)\Bigg(1 + \frac{1}{12i\mu}\Bigg)\\
&\simeq \frac{2}{\pi\mu}\Bigg(1 + \frac{i}{4\mu}({p'}^2-p^2)\eta^2\Bigg) \Bigg(\frac{p'}{p}\Bigg)^{i\mu}.
\end{align}

With the help of (\ref{hh}), we can now compute the squared absolute value of the temporal integral from the expression of the energy (\ref{dEdk}):
\begin{align}\label{integraleta}
  & \left\vert\int_0^\infty d\eta\, \eta \,\mathcal{H}_\nu^{(1)}(p'\eta)\mathcal{H}_\nu^{(2)}(p\eta)\,e^{-ik\eta-\epsilon\eta} \right\vert^2= \\
\nonumber    & =\left\vert\frac{2}{\pi\mu} \Bigg(\frac{\,p'}{p}\Bigg)^{i\mu}\Bigg[ \frac{1}{(ik+\epsilon)^2} + \frac{3}{2\mu}\frac{i({p'}^2 - p^2)}{\,(ik + \epsilon)^4}\Bigg] \right\vert^2 \\
\nonumber &=\frac{4}{\pi^2\mu^2} \Bigg( \frac{1}{(k^2 + \epsilon^2)^2}  + \frac{3k\epsilon}{\mu}\frac{{p'}^2 - p^2}{(k^2 + \epsilon^2)^4} + \frac{9}{4\mu^2}\frac{({p'}^2 - p^2)^2}{(k^2 + \epsilon^2)^4}\Bigg).
\end{align}

Gathering all terms we can now obtain via eq.(\ref{energia}) the energy emitted under a unit solid angle and frequency:

%\nonumber &=\frac{4}{\pi^2\mu^2} \Bigg( \frac{1}{(k^2 + \epsilon^2)^2} + \frac{9}{4\mu^2}\frac{({p'}^2 - p^2)^2}{(k^2 + \epsilon^2)^4} + \frac{3k\epsilon}{\mu}\frac{{p'}^2 - p^2}{(k^2 + \epsilon^2)^4}\Bigg)
\begin{align}
&\frac{dE}{dkd\Omega} = k^2\,\frac{e^2 {p'}^2\sin^2\theta}{2(2\pi)^3}\,\frac{1}{\mu^2} \Bigg( \frac{1}{(k^2 + \epsilon^2)^2}\\
\nonumber & - \frac{3k\epsilon}{\mu}\frac{(k^2 -2 {p'}^2k\cos{\theta})}{(k^2 + \epsilon^2)^4}+\frac{9}{4\mu^2}\frac{(k^2 -2 {p'}^2k\cos{\theta})^2}{(k^2 + \epsilon^2)^4} \Bigg),
\end{align} %poze Probabilitate
where the integration over the final momentum $p$ was rendered trivial due to the Dirac delta function ($p^2 = {p'}^2 + k^2 -2kp'\cos\theta$).  \par

We note that all integrals are of the following form:
\begin{equation}
\int dk \, k^2 \frac{k^\alpha}{(k^2 + \epsilon^2)^\beta} = \frac{\mathrm{\Gamma}( \frac{3 + \alpha}{2})\mathrm{\Gamma}( \frac{2\beta - \alpha - 3}{2})}{\epsilon^{2\beta - \alpha - 3}\,\mathrm{\Gamma}(\beta)}.
\end{equation}

This leads us to the resulted angular distribution of emitted energy:
\begin{align} \label{dEdOm}
\nonumber \frac{dE}{d\Omega} =& \,\frac{e^2 {p'}^2\sin^2\theta}{16\pi^2}\,\frac{1}{4\epsilon\mu^2}\Bigg\{ 1 - \frac{1}{\mu}\Big(\frac{2}{\pi} - \frac{3p'\cos{\theta}}{4\epsilon}\Big)\\
 &+ \frac{1}{\mu^2}\Big(\frac{45}{32} - \frac{6p'\cos{\theta}}{\pi\epsilon}+\frac{9{p'}^2\cos^2{\theta}}{8\epsilon^2\mu^2}\Big) \Bigg\}.
\end{align}

\begin{figure}
\centering
\includegraphics[width=2.8in]{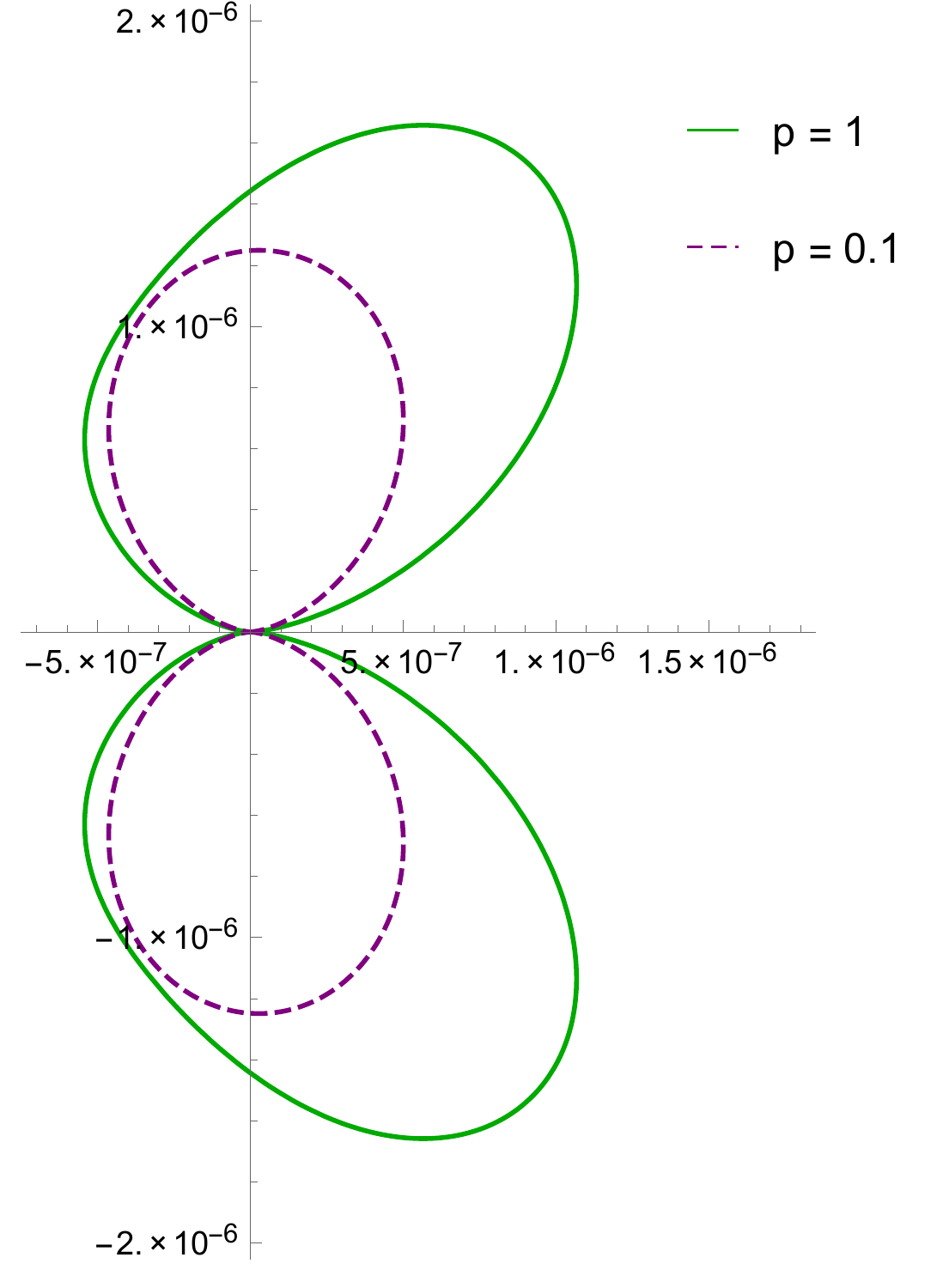}
\caption{\small Angular distribution of the emitted radiation, for $\mu = 100$. For small momentum we see the characteristic $\sin^2$ distribution. Increasing the momentum of the source causes the energy to be emitted in a cone in the forward direction. The cut-off parameter is $\epsilon = 10^{-2}$ and the small momentum curve was enhanced by a factor of $10^{2}$.}
\label{2}
\end{figure}

Plotting eq.(\ref{dEdOm}) we observe: a) the characteristic $\sin^2$ distribution for the radiation in the case of vanishing momentum of the source, b) as we increase the source momentum the radiation is emitted in a narrowing cone around the direction of  motion, c) increasing amount of radiation in the backward direction. In order for formula (\ref{dEdOm}) to remain valid, the $1/\mu$ corrections must remain small as compared to the leading term. We require thus that $p'/\epsilon \ll \mu$. \par
A further integration over $d\Omega$ gives us the total energy emitted in the process:
\begin{align}\label{Enonrel}
 E = \frac{e^2}{6\pi}\left(\frac{p'}{\mu}\right)^2\frac{1}{4\epsilon}\Bigg\{ 1 - \frac{2}{\pi\mu} + \frac{1}{\mu^2}\Big( \frac{45}{32} + \frac{9{p'}^2}{40 \epsilon^2} \Big) \Bigg\}
\end{align}
If we write $E = E_{cl} + E^{(1)}+E^{(2)}$ we can identify from eq.(\ref{Enonrel}) the lowest order term as
\begin{equation} \label{Ecl}
E_{cl} = \frac{e^2{p'}^2\omega^2}{6\pi m^2} \frac{1}{4\epsilon}.
\end{equation}
Guided by the results of Ref.\cite{N1}, we consider the acceleration to be
\begin{equation}
\ddot{x}(\eta) = \frac{d}{d\eta} \left(\frac{p}{p^0}\right).
\end{equation}
For our non-relativistic approximation this gives
\begin{eqnarray}
\ddot{x}(\eta) &\simeq& \frac{d}{d\eta} \left(\frac{p}{m} \omega \eta\right) \nonumber \\
 &=& \frac{1}{m} \frac{dp_{\text{phys}}}{d\eta} \nonumber \\
 &=& \frac{p\omega}{m}
\end{eqnarray}
The remaining factor of $\frac{1}{4\epsilon}$ is due to the presence of the adiabatic cut-off. When we take the limit $\epsilon \rightarrow 0$ the energy diverges. This can be understood as follows: the role of the cut-off is to decouple the fields and thus halt the interaction on time scales larger than $1/\epsilon$. When we take the vanishing limit this is equivalent to considering an infinite interaction time. Then, the energy radiated with a constant rate, under an infinite time, will be infinite. This also holds for a constantly accelerated charge in flat space. The results are consistent with that of Ref.\cite{Higuchi}. \par
Interestingly, if we naively take the non-relativistic limit in the results of Ref.\cite{N1} and also consider the adiabatic cut-off, we would obtain a result that is twice larger than (\ref{Ecl}). This is due to the fact that their calculation was tailored for a conformally flat spacetime with the conformal time ranging over the complete real axis. For the particular case of dS, this would mean the global de Sitter space. A similar situation was reported in Refs.\cite{C1,C2,C3} for Coulomb scattering in the expanding de Sitter space. For the expanding patch of dS, described by the line element (\ref{dS}), the calculation in Ref.\cite{N1} breaks down in eq.(30) where the boundary terms were neglected and in the subsequent integration over frequencies. If we were instead to consider the non-relativistic approximation $(z\ll1)$, by neglecting from the beginning quantities of order $ (p/p^0)^2 \simeq z^2$ and with the adiabatic cut-off, the results would be identical to ours. \par
The leading quantum correction to the emitted energy is
\begin{eqnarray}
\frac{E^{(1)}}{E_{cl}} = -\frac{2}{\pi}\frac{\omega}{m}.
\end{eqnarray}
\par
A negative quantum correction was also reported in all similar studies \cite{Higuchi,N3,N4,N1,N2}, for charges evolving in external electromagnetic and gravitational fields. The fact that the quantum effect suppresses the classical result thus seems to be a generic feature in such contexts.  In Refs. \cite{N3,N1,N2} it is noted that the quantum corrections arise due to a non-local integration in time over the classical trajectory. In our case the trajectory is fixed, with constant acceleration $\ddot{x}(\eta) = \frac{p\omega}{m}$ and the non-locality is implicit in the result. On the other hand, in Ref.\cite{N4}, the authors do not find the aforementioned nonlocality for the case of a charge moving in an electromagnetic plane-wave background. The difference is that this calculation is performed using the \emph{Schwinger-Keldysh }(\emph{in-in}) \emph{formalism}. It remains an open question why this difference arises. It will be an interesting subject for future work to calculate the radiation of a charge in de Sitter space using the in-in formalism and to compare with the results obtained in this paper . \par
In Ref.\cite{N2} it is found that the first correction to the radiation of a charge moving in a conformally flat background contains third derivative terms. Up to the orders that we have considered in our case we have $\dddot{x\,}(\eta) \simeq  0$. The fact that we have a non-zero first order contribution thus suggests that our method captures terms that the WKB approximation misses.

\subsection{Radiation in the ultra-relativistic limit}

In this section we examine the behavior of the probability and the emitted energy through the process in the ultra-relativistic limit. Starting from (\ref{masterformula}) and imposing the condition for ultra-relativistic motion of the source ($z\gg 1$), we obtain:
\begin{equation}
I_{p,p'}^\nu(\eta) = \frac{2}{\pi\eta\sqrt{p'p}} e^{i\eta(p'-p)}.
\end{equation}
The energy radiated under a unit solid angle and in unit frequency thus becomes:
\begin{align}
\nonumber E &= \frac{e^2\pi^2}{4}\frac{{p'}^2\sin^2\theta}{2(2\pi)^3}\left\vert\frac{2}{\pi\sqrt{p'p}}\int_0^\infty d\eta \, e^{i\eta(p' - p - k + i\epsilon)} \right\vert^2\\
&= \frac{e^2}{2(2\pi)^3}\frac{\,p'}{p} \frac{1}{(p' - p - k)^2 + \epsilon^2}.
\end{align}
Integrating over the momenta of the photon we obtain the total energy emitted in the process:
\begin{equation}\label{integral_ur}
E = \frac{e^2}{8\pi^2}\int_0^\infty k^2 dk \int_{-1}^{1} d(\cos{\theta})\frac{{\,p'}}{p}\frac{\sin^2{\theta}}{(p' - p - k)^2 + \epsilon^2}.
\end{equation}

By changing the integration variable to \\
$p = \sqrt{{p'}^2 + k^2 - 2p'k\cos{\theta}}$, the angular integral becomes:
\begin{equation}
E = \frac{e^2}{8\pi^2}\int_0^\infty k\,dk\int_{|p' - k|}^{p' + k} dp\, \frac{1 - \frac{(p^2 - {p'}^2 - k^2)^2}{4{p'}^2k^2}}{(p' - p - k)^2 + \epsilon^2}.
\end{equation}

A further change of variable to $z = p - p' + k$ results in:
\begin{equation}
E = \frac{e^2}{8\pi^2}\int_0^\infty k\,dk\int_{|p' - k| - p' + k}^{2 k} dz \frac{1 - \frac{(z^2 - 2z(p' - k) - 2kp')^2}{4{p'}^2k^2}}{z^2 + \epsilon^2}.
\end{equation}

The indefinite integral over $z$ has the following result:
\begin{align}
\nonumber \mathcal{B}(z)=&-\frac{1}{4k^2{p'}^2}\Big\{ \frac{1}{3}z(12k^2 + 12 {p'}^2 + 6p'z + z^2 -6k(6p'
\\+& z)
\nonumber - 3\epsilon^2) + \epsilon(-4k^2 + 12kp' - 4{p'}^2 + \epsilon^2)\arctan{\frac{z}{\epsilon}} \\
 +& 2(2k^2p' - p'\epsilon^2 + k( -2{p'}^2+\epsilon^2))\log{(z^2+\epsilon^2)}\Big\}
\end{align}

Using the notation introduced above we can write the energy emitted in the process as:
\begin{align} \label{Erel}
E =& \frac{e^2}{8\pi^2}\int_0^{p'} k\,dk\, \Big[\mathcal{B}(2k) - \mathcal{B}(0)\Big] \nonumber\\
&+ \frac{e^2}{8\pi^2} \int_{p'}^{\infty} k\,dk\, \Big[\mathcal{B}(2k) - \mathcal{B}(2k-2p')\Big]
\end{align}

\begin{figure}
\centering
\includegraphics[width=3.8  in]{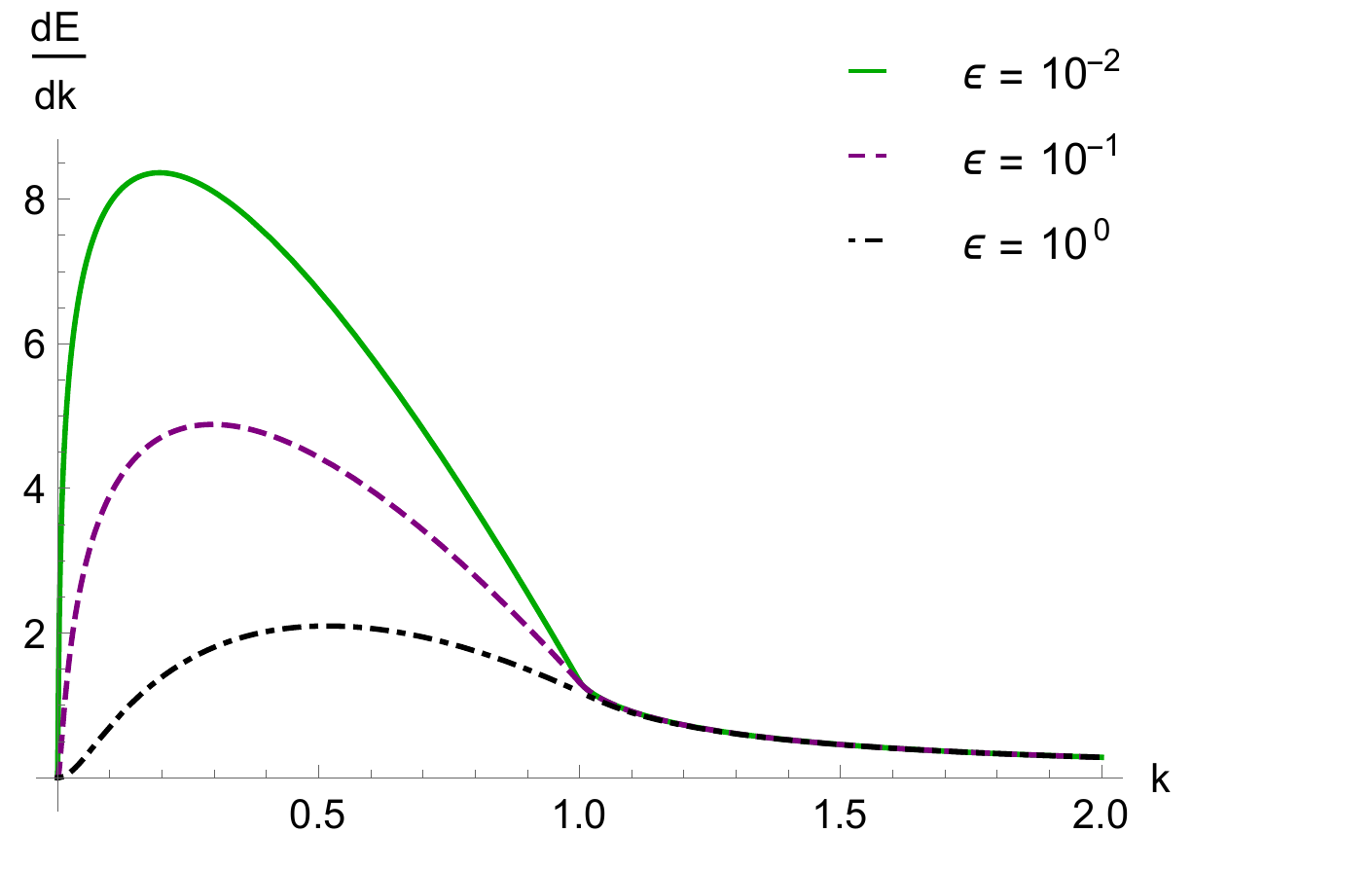}
\caption{\small Frequency distribution of the radiated energy in the ultra-relativistic limit for $p'=1$. For large frequencies the radiation falls off as 1/k.}
\label{3}
\end{figure}

In fig.(\ref{3}) we have plotted the integrand of (\ref{Erel}), which is the frequency distribution of the energy. The bulk of the radiation is emitted under frequencies $k \leq p'$ as one would expect on physical grounds. For a small cut-off we see that most of the radiation is emitted for small frequencies. Increasing the $\epsilon$ parameter reveals that there is actually another competing channel around $k\simeq p'$. This is not present in the non-relativistic case. We can understand this as follows: because we are investigating the process under weak gravitational field conditions, there is a loose energy conservation principle at action, which is reminiscent from flat space. For the non-relativistic case, where the energies go as $\sim p^2$, the photon momentum cannot compete with the source, and thus the only route towards energy conservation is $p\simeq p', k \rightarrow 0$. On the other hand in the ultra-relativistic limit, because the energies go as $\sim p$, the energy of the photon  is on the same footing as the energy of the source, and the channel with $k \simeq p', p\rightarrow 0$ becomes relevant. Thus we understand the peak at $k \simeq p'$ as arising from an interplay between the gravitational field, which gently lifts the energy conservation constraint, and the relativistic regime, which puts the energy of the radiation on a par with that of the source.

For large frequencies $k > p'$ we have a tail that falls-off as $1/k$, which leads to a logarithmic divergence when integrated over. The presence of this divergence is intimately linked to the famous divergence problem of de Sitter space \cite{Ak1}. We can understand it as a symptom of the finite integration over conformal time in (\ref{dEdk}). Because the Maxwell field is conformal, the photon effectively "lives" in conformal time and "feels" the limit $\eta \rightarrow 0$ as being abrupt, although in the physical picture everything seems to be diluted away smoothly by the expansion of space. The finite limit for the temporal integration manifests like a finite-time sudden cut-off which leads to transitory effects, undesirable divergences and other artifacts \cite{Ak6,Ak3,Ak4,Ak5,Nk}.

\section{Discussion}
In this paper we have analysed the quantum radiation of a charge evolving on the expanding de Sitter spacetime. The emitted energy was derived from the transition amplitude of the corresponding sQED process. We compared the results of our perturbative calculation with that of Ref.\cite{N1}, which was done in the WKB approximation.
We have obtained the radiated energy as a power series in the Hubble constant, in the asymptotic case of a weak gravitational field. For a non-relativistic motion of the source, we have found the leading term to be compatible with the expected classical results. This is also identical to the results of Ref.\cite{N1}, within the same approximation. Furthermore, the leading quantum correction was found to be negative, a result also reported in all similar studies.  In the ultra-relativistic limit we expected to obtain a result which takes the form of the relativistic generalization of the Larmor formula. Instead we found that the energy has a logarithmic divergence for large frequencies. We interpret this as follows: the finite integration limit for the conformal time mimics a sudden decoupling of the interaction at time $\eta \rightarrow 0$. Because this "event" happens under an arbitrarily small time interval, arbitrarily high frequency modes can get excited. Thus we also understand why this effect does not show up in the non-relativistic case, where only small frequency photons are emitted. \par
It would be interesting to see whether the above mentioned pathological fingerprint also shows up in a classical calculation, for the same setup.  There are a number of papers that deal with the radiation of classical charges evolving on the global dS \cite{MT,Bicak1,Bicak2,Bicak3}. For the expanding patch of dS, the only study that we are aware of is done in Ref.\cite{Ak2}. We note that our results are compatible with that of Ref.\cite{Ak2}, in that we find that comoving observers see no radiation. Indeed if we set $p'=0$ in eq.(\ref{Enonrel}) and eq.(\ref{Erel}) we find vanishing energy in both non-relativistic and relativistic cases. The situation is similar to the uniformly accelerated case in flat space. It was shown in Ref.\cite{Kalinov15} that if we consider the problem in a non-inertial (Rindler) reference frame: while the observers which are co-accelerated with the charge see no radiation, if there is mutual motion between the observer and the charge in the Rindler frame, an energy-flux will be present. It would be interesting to do a systematic study in the lines of Ref.\cite{Kalinov15}, of the classical radiation emitted by charges on arbitrary trajectories on the expanding dS. Also it would be interesting to see how our results change if we consider proper Dirac electrons. \par
One more thing is worth noting. It is a pleasing fact that out of all 1st order processes the one studied here is the only one that falls off as an inverse power of $\mu$ as we go towards the flat space limit. As we have also signaled in Ref.\cite{B1}, the probabilities for all other 1st order processes are exponentially suppressed as $e^{-\alpha(\theta)\mu}$, including the one depicted in Fig.(\ref{1b}). This is linked to the fact that the process in Fig.(\ref{1a}), that forms the object of this study, is the only one that has a classical analogue.

\begin{acknowledgements}
We would like to thank Ion I. Cot\u aescu, Cosmin Crucean, Nistor Nicolaevici and Victor E. Ambru\c s for fruitful discussions that greatly enhanced the quality of this work.
We sincerely thank the two anonymous reviewers for their insightful comments that helped improve and clarify this manuscript.
\end{acknowledgements}

\end{document}